\documentclass[pra,twocolumn,preprintnumbers,floatfix]{revtex4}
\usepackage{graphicx}
\usepackage{dcolumn}
\usepackage{bm}
\usepackage{latexsym,epsfig}
\usepackage{graphicx}
\usepackage{verbatim}
\usepackage{comment}
\usepackage{amsmath}
\usepackage{amssymb}
\usepackage{stmaryrd}
\usepackage{color}
\usepackage{epstopdf}
\usepackage{grffile}
\usepackage{float}
\usepackage[normalem]{ulem}
\DeclareGraphicsExtensions{.eps}

\newcommand{\beq}{\begin{equation}}
\newcommand{\eeq}{\end{equation}}
\newcommand{\bea}{\begin{eqnarray}}
\newcommand{\eea}{\end{eqnarray}}
\newcommand{\ben}{\begin{eqnarray*}}
\newcommand{\een}{\end{eqnarray*}}
\newcommand{\bfig}{\begin{figure}}
\newcommand{\efig}{\end{figure}}

\definecolor{BLUE}{rgb}{0,0,1} 

\usepackage{placeins}
\usepackage{afterpage}
\usepackage{multirow}
\begin{document}
\title{Probing Deconfined Quantum Criticality Point on a Digital Quantum Computer}
\author{Mrinal Kanti Giri}
\email{mrinalphy333@gmail.com}
\author{Pinaki Sengupta}
\email{psengupta@ntu.edu.sg}
\affiliation{School of Physical and Mathematical Sciences, Nanyang Technological University, Singapore 637371}

\date{\today}

\begin{abstract}
A deconfined quantum critical point (DQCP) is a continuous transition between ordered phases with unrelated broken symmetries, beyond the Landau-Ginzburg-Wilson paradigm. In one dimension, such a transition between a $z$-polarized ferromagnet (zFM) and a valence-bond solid (VBS) is well established numerically in a frustrated spin-$1/2$ chain. Here we ask whether a superconducting quantum processor can resolve this transition and its signatures despite finite-size effects and device noise. The ground states are prepared by ancilla-projected quantum imaginary-time evolution (QITE), whose exact target states are compressed by tensor-network optimization into shallow depth quantum circuits. The hardware data distinguish the phases through their order parameters, spin and dimer correlations, structure factors, correlation ratios, and snapshot distributions. The crossing of the order parameters locates the transition, and its finite-size extrapolation lies close to the thermodynamic-limit value obtained from large-scale numerics. A noise model based on the device calibration accounts quantitatively for the reduced amplitudes measured on hardware. Our findings demonstrate that a Landau-forbidden transition between competing orders can be resolved on current quantum processors.
\end{abstract}

\maketitle

\section{Introduction}

Continuous phase transitions are conventionally described within the Landau-Ginzburg-Wilson (LGW) paradigm, in which a single order parameter distinguishes the two phases and the ordered phase breaks a subgroup of the symmetry of the disordered one. A direct continuous transition between two phases that break unrelated symmetries is then generically forbidden: it should either be first order or proceed through an intermediate phase. Deconfined quantum critical points (DQCPs) evade this constraint~\cite{Senthil2004Science,Senthil2004PRB}. At a DQCP, the two order parameters are not independent but arise from different confinements of fractionalized degrees of freedom, that become deconfined at the transition, so that the suppression of one order and the emergence of the other are two aspects of the same critical point. The canonical example, the transition between a N\'eel antiferromagnet and a valence-bond solid in two dimensions, has been studied intensively using both analytical methods and numerical simulations of designer microscopic Hamiltonians.~\cite{Sandvik2007} While its existence in the current 2D microscopic models has come under debate,~\cite{Cui2025} DQCP is well established in 1D spin systems. 

A particularly clean incarnation of this physics was identified in one dimension~\cite{JiangMotrunich2019,Roberts2019,Huang2019,Mudry2019}. In a spin-$1/2$ chain with $\mathbb{Z}_2\times\mathbb{Z}_2$ symmetry, the competition between ferromagnetic nearest-neighbor and antiferromagnetic next-nearest-neighbor interactions drives a direct transition between a $z$-polarized ferromagnet (zFM), which breaks a spin-flip symmetry, and a valence-bond solid (VBS), which breaks lattice translation. The transition is described by a Luttinger-liquid theory in which both order parameters are controlled by the same critical field, and it displays DQCP phenomenology, including an emergent enlarged symmetry and equal scaling of the two orders at criticality~\cite{Roberts2019,Huang2019,Xi2022}. Because the transition point is known accurately from large-scale matrix-product-state calculations~\cite{Roberts2019}, this chain provides a well-controlled benchmark for exploring Landau-forbidden criticality with quantum simulators.

Programmable quantum devices offer a new route to such problems. Analog Rydberg-atom arrays have resolved quantum phases and critical dynamics in large systems~\cite{Keesling2019,Ebadi2021}, and realizations of Landau-forbidden criticality in such arrays have been proposed~\cite{Lee2023,Bombieri2025}. On digital processors, phase transitions have so far been studied mainly with states of known structure~\cite{Satzinger2021,Smith2022} or in small systems~\cite{Kottmann2021,Herrmann2022}. The central obstacle on current noisy intermediate-scale devices~\cite{Preskill2018} is ground-state preparation: for frustrated Hamiltonians with competing orders, imaginary-time and variational approaches typically require circuits deeper than current hardware allows~\cite{Motta2020,McArdle2019,Ejima2025,Yi2025}, and noise can destroy the targeted phase~\cite{Khindanov2025}. To our knowledge, a direct transition between two distinct symmetry-broken orders has not yet been resolved on a digital quantum processor.

In this work we ask whether a digital quantum processor can resolve the deconfined quantum criticality and its signatures in the model mentioned earlier. We prepare the ground state by ancilla-projected quantum imaginary-time evolution, compressed by tensor-network optimization~\cite{Lin2021} into shallow circuits that match the device connectivity, and measure the two order parameters together with the spin and dimer correlations on both sides of the transition. The hardware resolves both ordered phases, and the crossing of the order parameters locates the transition in agreement with noiseless simulations and large-scale numerics. Correlation ratios, which characterize the sharpness of the structure-factor peaks and are largely insensitive to the overall attenuation of the signal by noise, show on hardware how the two orders exchange their roles across the transition. A noise model based on the device calibration reproduces the data and identifies the main limitations of the approach; compiling the circuits against the exact projection probability of the target state mitigates the most severe of them.

The paper is organized as follows. Section~\ref{sec:model} introduces the model and the order parameters. Section~\ref{sec:methods} describes the ancilla-projected QITE, the circuit compression and the hardware implementation, and defines the measured correlation functions and structure factors. Section~\ref{sec:results} presents the hardware results, the correlation ratios, and the noise model, and Sec.~\ref{sec:conclusion} summarizes our findings. Details of the noise simulations are given in Appendix~\ref{app:noise}, and the role of the projection probability and the $p_0$-aware compilation in Appendix~\ref{app:p0_draft}.

\section{Model}
\label{sec:model}
We consider a one-dimensional spin-$1/2$ chain with nearest-neighbour (NN) ferromagnetic and next-nearest-neighbour (NNN) anti-ferromagnetic interactions described by the Hamiltonian
\begin{align}
    H_0 =&  \sum_{i} -J_x \sigma^x_i \sigma^x_{i+1} - J_z \sigma^z_i \sigma^z_{i+1} \notag \\ 
    &+ \sum_{i} K_x \sigma^x_i \sigma^x_{i+2} + K_z \sigma^z_i \sigma^z_{i+2}, 
    \label{eq:dqcp_model}
\end{align}
where $\sigma_i^\alpha$ $(\alpha=x,y,z)$ are Pauli matrices acting on site $i$. $J_x,J_z>0$ define ferromagnetic NN interactions, while $K_x,K_z>0$ denote antiferromagnetic NNN interactions. The competition between these two terms frustrates simple magnetic ordering and stabilizes distinct ordered phases - providing a minimal setting for realizing competing magnetic and VBS orders. 

The Hamiltonian $H_0$ is invariant under the two Ising-like spin-flip symmetries $g_x = \prod_i \sigma^x_i, \; g_z = \prod_i \sigma^z_i$, which together generate $\mathbb{Z}_2\times\mathbb{Z}_2$ (the two generators commute for even chain length $L$, used throughout this work), as well as translation, spatial inversion, and time reversal symmetries. 
When $J_z>J_x$ the $z$-channel FM coupling dominates and the ground state develops ferromagnetic order along $z$, denoted zFM. This phase spontaneously breaks $g_x$, since $g_x\,\sigma^z_i\,g_x = -\sigma^z_i$. We characterize it by the uniform magnetization
\begin{equation}
M^{\mathrm{FM}}_z = \frac{1}{L}\sum_{j=1}^{L}\langle\sigma^z_j\rangle .
\label{eq:Mz}
\end{equation}
Conversely, for $J_x>J_z$ the system favours an $x$-polarized phase (xFM) that breaks $g_z$. In the following we focus on the zFM--VBS transition.
\begin{figure}[t]
    \centering
    \includegraphics[width=1\columnwidth]{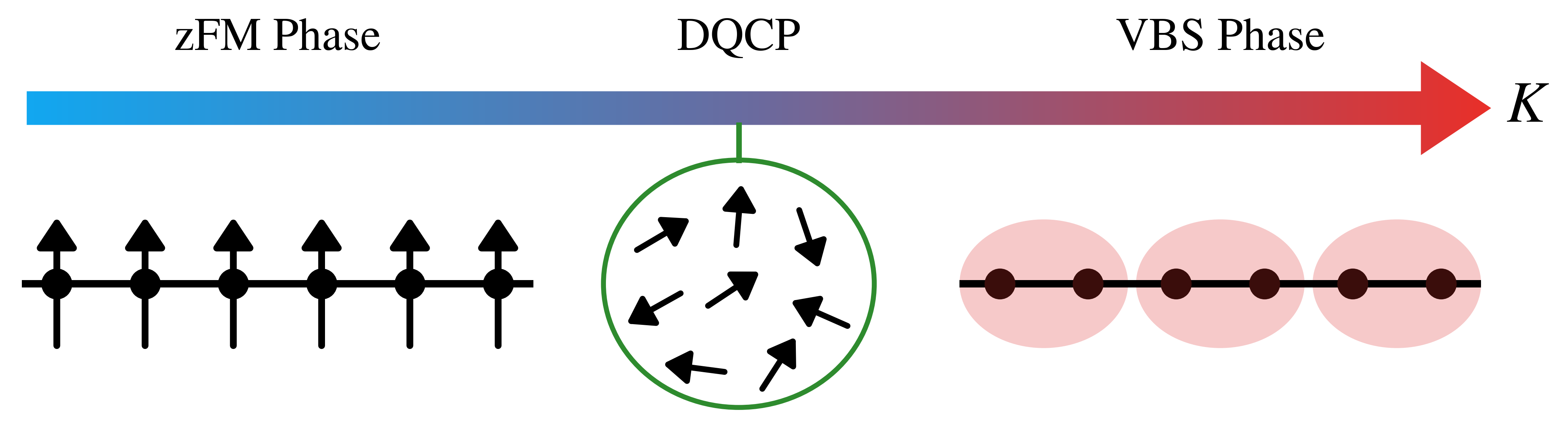}
    \caption{Schematic of the zFM--VBS transition. With increasing frustration $K$, the $z$-polarized ferromagnet (zFM, left), which breaks the spin-flip symmetry, gives way to a valence-bond solid (VBS, right) with dimerized bonds on alternating links, which breaks translation symmetry.}
    \label{fig:cartoon}
\end{figure}
For sufficiently strong, nearly isotropic NNN interactions the magnetic order is suppressed and the system enters a VBS phase in which the spins dimerize on alternating bonds, spontaneously breaking translation symmetry. We define the local bond operator
\begin{equation}
B_j = \sigma^x_j\sigma^x_{j+1} + \sigma^z_j\sigma^z_{j+1},
\label{eq:Bj}
\end{equation}
and the VBS order parameter as the staggered bond modulation
\begin{equation}
\Psi_{\mathrm{VBS}} = \frac{1}{L}\sum_{j=1}^{L}(-1)^j\,\langle B_j\rangle .
\label{eq:PsiVBS}
\end{equation}

On a finite chain with an exactly symmetric ground state the linear order parameters in Eqs.~\ref{eq:Mz} and \ref{eq:PsiVBS} vanish identically. To expose the broken-symmetry sector we add small symmetry-conjugate probe fields,
\begin{equation}
H = H_0 \;-\; h_z\sum_i \sigma^z_i
        \;-\; h_d\sum_i (-1)^i B_i ,
\label{eq:Hprobe}
\end{equation}
where $h_z$ and $h_d$ are small fields conjugate to the zFM and VBS order parameters: $h_z$ breaks $g_x$ to render $M^{\mathrm{FM}}_z$ nonzero, and $h_d$ breaks one-site translation to render $\Psi_{\mathrm{VBS}}$ nonzero. The two fields are never applied simultaneously. For each $K$ we prepare two separate ground states: one with $(h_z,h_d)=(0.003,0)$, which selects the zFM sector and is used for $M^{\mathrm{FM}}_z$ and all spin observables, and one with $(h_z,h_d)=(0,0.003)$, which selects the VBS sector and is used for $\Psi_{\mathrm{VBS}}$ and all dimer observables. The field strength $0.003$ is small enough to select the respective broken-symmetry sectors without appreciably shifting the critical point. Varying the field strength between $0.001$ and $0.01$ shifts $K_c(L)$ by less than $0.001$ (see Appendix~\ref{app:validation}); only for $L=8$ does $h=0.003$ select the broken-symmetry states partially.

In this study, we adopt the following parametrization:
\begin{equation}
K_x = K_z \equiv K, \qquad
J_z = J(1+\delta), \qquad
J_x = J(1-\delta),
\label{eq:param}
\end{equation}
and set $J=1$ as the energy scale. The anisotropy $\delta>0$ biases the NN coupling towards the $z$ direction (favouring zFM), while $K$ tunes the frustrating NNN interaction. We fix $\delta = 0.5$ and sweep $K$ across the range $0.40 \le K \le 0.65$, which crosses the zFM-VBS phase boundary of Refs.~\cite{Roberts2019,Huang2019}. 

We have used an IBM quantum processor (IBMQ) to probe this non-Landau continuous transition between two incompatible ordered phases. We employ quantum imaginary-time evolution to prepare the low-energy ground state of Eq.~\ref{eq:Hprobe} and probe the transition through the magnetic order parameter, the VBS order parameter, correlation functions and structure factors. The details are given in Sec.~\ref{sec:methods}.

\section{Methods}
\label{sec:methods}
\subsection{Ground-state preparation using ancilla-projected QITE}
\begin{figure*}[t]
    \centering
    \includegraphics[width=\textwidth]{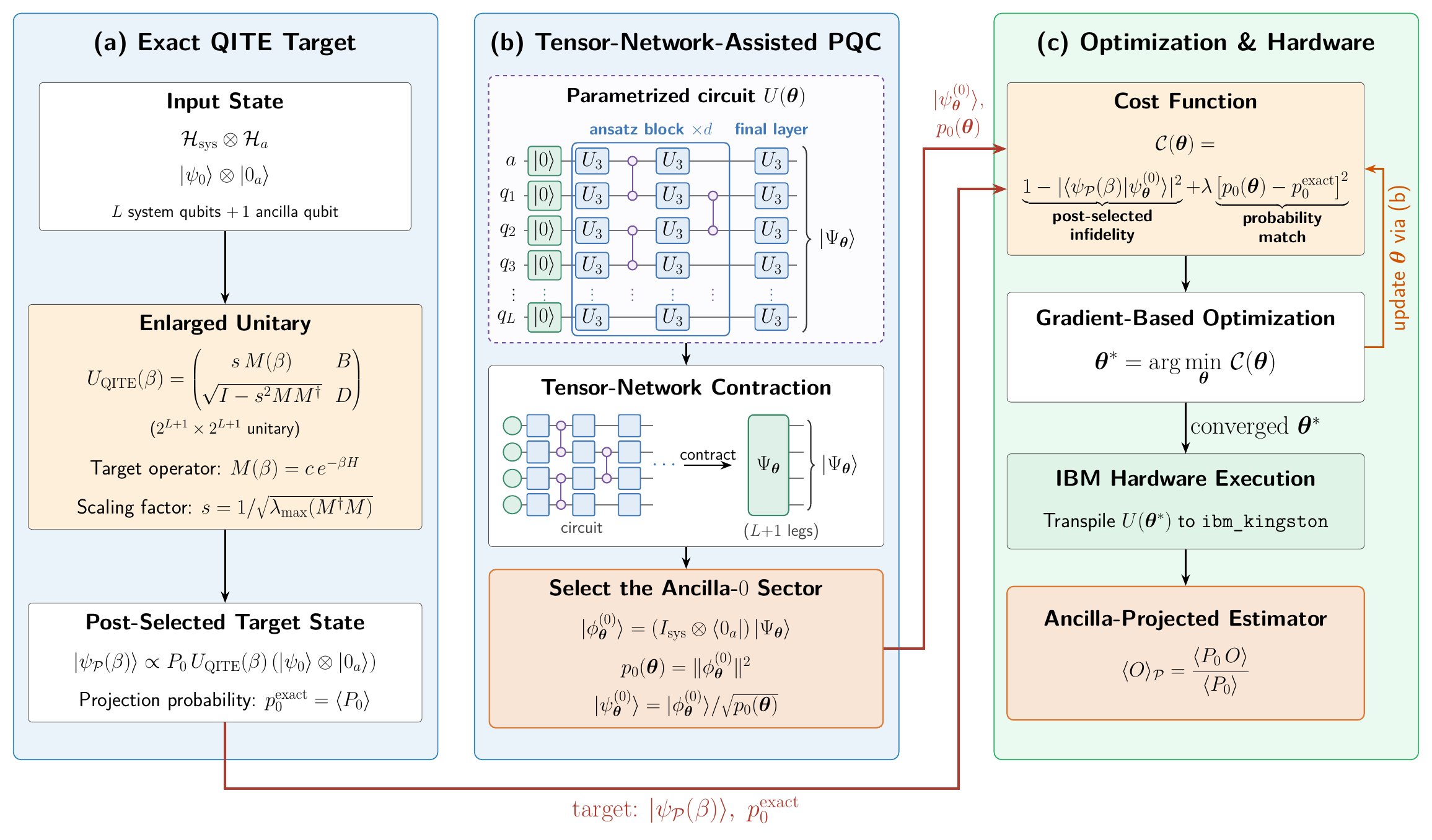}
    \caption{Workflow of the variational post-selected QITE scheme. (a) Exact QITE target: the non-unitary propagator $M(\beta)=c\,e^{-\beta H}$ is embedded, with scaling factor $s$, in an enlarged unitary $U_{\rm QITE}(\beta)$ acting on $L$ system qubits and one ancilla. Projecting the output onto the ancilla-$|0_a\rangle$ branch gives the target state $|\psi_{\mathcal P}(\beta)\rangle$ and its projection probability $p_0^{\rm exact}$. (b) Tensor-network-assisted PQC: the brickwork ansatz of $U_3$ rotations and nearest-neighbour CZ gates on even and odd bonds, repeated $d$ times and followed by a final $U_3$ layer, is represented as a network of gate tensors and contracted exactly into a single tensor $\Psi_{\bm\theta}$ with $L+1$ open legs, i.e., the output state $|\Psi_{\bm\theta}\rangle$. Selecting the ancilla-$0$ sector gives $|\psi^{(0)}_{\bm\theta}\rangle$ and $p_0(\bm\theta)$. (c) Optimization and hardware: the cost function of Eq.~\ref{eq:cost} is minimized with JAX automatic differentiation and Adam, with each parameter update re-contracted in (b). The converged circuit $U(\bm\theta^*)$ is transpiled to and executed on \texttt{ibm\_kingston}. All observables are evaluated with the ancilla-projected estimator of Eq.~\ref{eq:meas}.}
    \label{fig:workflow}
\end{figure*}
In this section, we present an ancilla-projected QITE framework for ground-state preparation and observable measurement on digital quantum computer.
For a given Hamiltonian $H$, an approximate ground state can be prepared by applying the imaginary-time evolution operator $e^{-\beta H}$ to an initial state $|\psi_0\rangle$. The normalized imaginary-time-evolved state is 
\begin{equation}
|\psi(\beta)\rangle =
\frac{e^{-\beta H}|\psi_0\rangle}
{\sqrt{\langle \psi_0|e^{-2\beta H}|\psi_0\rangle}},
\end{equation}
where $\beta$ is the imaginary time. For sufficiently large $\beta$, the excited-state components are exponentially suppressed, and $|\psi(\beta)\rangle$ approaches the ground state of the system. Since $e^{-\beta H}$ is non-unitary, it cannot be implemented directly as a gate operation on a digital quantum computer. In local-unitary QITE~\cite{Motta2020}, each Trotter step is replaced by a unitary on a domain of $D$ qubits, determined from expectation values measured in that domain. Since $D$ must cover the correlation length, which at a transition grows with the system size, and the measurement and gate costs grow exponentially with $D$ while the circuit depth grows linearly with $\beta$~\cite{Motta2020,Nishi2021}.
In contrast, we use an ancilla-projected formulation, where the non-unitary imaginary-time propagator is embedded into a larger unitary $U_{\rm QITE}(\beta)$ acting on both the system and single ancilla qubits~\cite{Ejima2025,Yi2025}. By applying the enlarged unitary $U_{\rm QITE}(\beta)$ to the initial system - ancilla state $|\psi_0\rangle\otimes |0_a\rangle$, the output state takes the form as -
\begin{equation}
U_{\rm QITE}(\beta)
\left(|\psi_0\rangle \otimes |0_a\rangle\right) =
c e^{-\beta H}|\psi_0\rangle \otimes |0_a\rangle + |\Phi_\perp\rangle .
\end{equation}
Here, first term represents the imaginary-time-evolved component associated with the ancilla-$|0_a\rangle$ branch. The state $|\Phi_\perp\rangle$ contains all remaining components, and $c$ is a normalization factor. To extract the physical QITE state, we project onto this ancilla-$|0_a\rangle$ branch using $P_0 = I_{\rm sys}\otimes |0_a\rangle\langle 0_a|$. The normalized projected QITE state is then
\begin{equation}
|\psi_{\mathcal{P}}(\beta)\rangle =
\frac{ P_0 U_{\rm QITE}(\beta) \left(|\psi_0\rangle \otimes |0_a\rangle\right)} {\sqrt{p_0}},
\label{eq:proj}
\end{equation}
where
\begin{equation}
p_0 =\langle \psi_0,0_a|U_{\rm QITE}^{\dagger}(\beta) P_0 U_{\rm QITE}(\beta) |\psi_0,0_a\rangle
\end{equation}
is the corresponding projection probability. For a system observable $O$, the relevant expectation value must be evaluated within the successful ancilla-$|0_a\rangle$ branch. This is represented by the conditional expectation value
\begin{equation}
\langle O\rangle_{\mathcal{P}} =
\frac{\langle \Psi| P_0 (O\otimes I_a) P_0|\Psi\rangle}{\langle \Psi|P_0|\Psi\rangle},
\end{equation}
where
\begin{equation}
|\Psi\rangle =
U_{\rm QITE}(\beta)
\left(|\psi_0\rangle \otimes |0_a\rangle\right).
\end{equation}
Since $O$ acts only on the system qubits and commutes with $P_0$, this reduces to
\begin{equation}
\langle O\rangle_{\mathcal{P}} = \frac{\langle P_0 O\rangle}{\langle P_0\rangle}.
\label{eq:meas}
\end{equation}
A larger imaginary time $\beta$ improves convergence to the ground state; we fix $\beta=100$ throughout, for which the crossing points $K_c(L)$ agree with those at $\beta=400$ to better than $10^{-5}$ (see Appendix~\ref{app:validation}).

\subsection{Parametrized quantum circuit (PQC) and IBMQ implementation}

For each value of the tuning parameter $K$, we construct a hardware-efficient parametrized quantum circuit (PQC) acting on the $L$ system qubits and one ancilla qubit to represent the ancilla-projected QITE evolution. The circuit consists of single-qubit rotation gates and two-qubit entangling gates arranged according to the connectivity of the target IBM Quantum backend. Implementing the full enlarged unitary $U_{\rm QITE}(\beta)$ as a gate-level circuit is not suitable for present hardware: $U_{\rm QITE}(\beta)$ is a dense $2^{L+1}\times 2^{L+1}$ unitary, and the exact decomposition of a generic $n$-qubit unitary requires a number of two-qubit gates that grows exponentially, as $\mathcal{O}(4^{n})$, with $n$. Instead, we map the target QITE state onto the chosen compact PQC ansatz.

The ansatz is a brickwork circuit of depth $d$, as shown in Fig.~\ref{fig:workflow}(b). Each of the $d$ layers consists of general single-qubit rotations $U_3$ on all qubits followed by CZ gates on even bonds, and a second set of $U_3$ rotations followed by CZ gates on odd bonds; a final layer of $U_3$ rotations completes the circuit. The CZ gates act only between nearest neighbours of an open chain of $L+1$ qubits, which maps directly onto a path of the heavy-hex lattice. We use $d=6,7,8,9$ for $L=8,10,12,14$, respectively, so that the circuits contain $d\times L=48,70,96,126$ CZ gates ($L$ per layer) with a CZ depth of $2d=12$--$18$ (even and odd bonds in each layer).

Note that the target states are obtained from the Hamiltonian of Eq.~\ref{eq:Hprobe} with periodic boundary conditions, whereas the ansatz itself has open-boundary connectivity; the correlations across the boundary are generated indirectly through the depth of the circuit, whose light cone ($2d$ CZ rounds) spans the entire chain for all depths used here. 

The circuit parameters $\bm\theta$ are optimized classically with a tensor-network method. We represent the circuit as a network of gate tensors in quimb~\cite{Gray2018}, contract it exactly to obtain the full output state $|\Psi_{\bm\theta}\rangle$ of the $L+1$ qubits, and compute the gradient of the cost with JAX automatic differentiation. The cost combines the post-selected infidelity to the target QITE state $|\psi_\mathcal{P}(\beta)\rangle$ of Eq.~\ref{eq:proj} with a penalty that matches the ansatz success probability to its exact QITE value,
\begin{equation}
\mathcal{C}(\bm\theta) =
\bigl(1 - |\langle\psi_\mathcal{P}(\beta)\,|\,\psi^{(0)}_{\bm\theta}\rangle|^2\bigr)
\;+\;
\lambda\,\bigl(p_0(\bm\theta) - p_0^{\mathrm{exact}}\bigr)^2 ,
\label{eq:cost}
\end{equation}
where $|\psi^{(0)}_{\bm\theta}\rangle$ is the renormalized $|0_a\rangle$ component of the circuit output, $p_0(\bm\theta)=\langle P_0\rangle$ is the ansatz success probability, $p_0^{\mathrm{exact}}$ is that of the exact-QITE target, and $\lambda=2$. We minimize $\mathcal{C}(\bm\theta)$ with the Adam optimizer~\cite{Kingma2014} for $3\times10^{4}$ iterations. The couplings $K$ are optimized in sequence, each warm-started from the converged parameters at the preceding $K$, and the first point from random angles. 
For small values of exact projection probability, $p_0^{\mathrm{exact}}<10^{-3}$, the penalty term in Eq.~\ref{eq:cost} is omitted and only the post-selected infidelity is minimized. The fidelities $\mathcal{F}=|\langle\psi_\mathcal{P}|\psi^{(0)}_{\bm\theta}\rangle|^2$ of the optimized circuits with their targets are given in Appendix~\ref{app:validation}. The complete workflow is summarized in Fig.~\ref{fig:workflow}.

After optimization the PQCs are transpiled to the native gate set $\{\mathrm{CZ}, R_z, \sqrt{X}\}$ of \texttt{ibm\_kingston}, a 156-qubit IBM Heron processor with heavy-hex connectivity. The executed circuits have total depths of $65$, $75$, $85$, and $95$ for $L=8$, $10$, $12$, and $14$, respectively, with the CZ counts unchanged from the logical circuits. Expectation values are evaluated with the IBM Quantum Runtime EstimatorV2 primitive. We employ twirled readout error extinction (TREX)~\cite{vandenBerg2022} to mitigate readout errors, and gate twirling~\cite{Wallman2016} over $32$ randomizations to convert coherent errors into stochastic Pauli noise. We set the target precision to $0.01$ per expectation value, corresponding to a sampling scale of order $1/(0.01)^2\sim10^4$ shots, which balances statistical accuracy against total hardware cost for the full parameter scan.

\subsection{Order parameters, correlation functions, structure factors, and snapshot distributions}
On IBMQ, all observables are evaluated through the ancilla-projected expectation
value $\langle\cdots\rangle_{\mathcal P}$ of Eq.~\ref{eq:meas}. We measure the zFM and VBS order parameters $M_z^{\rm FM}$ and $\Psi_{\rm VBS}$, defined in Eqs.~\ref{eq:Mz} and \ref{eq:PsiVBS}, to identify the phase transition. To further characterize the states, we compute the spin-spin and dimer-dimer correlation functions and the corresponding structure factors. The spin-spin correlation is
\begin{equation}
C_z(r) = \langle \sigma_0^z \sigma_r^z\rangle_{\mathcal P},
\qquad
S_z(q) = \sum_{r=0}^{L-1} \cos(qr)\,C_z(r),
\end{equation}
and the dimer-dimer correlation function is
\begin{equation}
C_D(r) = \langle B_0 B_r\rangle_{\mathcal P},
\qquad
S_D(q) = \sum_{r=0}^{L-1} \cos(qr)\,C_D(r),
\end{equation}
with $S_z(q)$ and $S_D(q)$ the associated structure factors, evaluated at $q=2\pi n/L$ ($n=0,\dots,L-1$). 
The sharpness of these peaks is quantified by the correlation ratio~\cite{Kaul2015,Pujari2016,Nomura2021}
\begin{equation}
R = 1-\frac{S(q_0+\Delta q)}{S(q_0)},\qquad \Delta q=\frac{2\pi}{L},
\label{eq:ratio}
\end{equation}
with $q_0=0$ for $S_z$ ($R_{\rm zFM}$) and $q_0=\pi$ for $S_D$ ($R_{\rm VBS}$). In the presence of long-range order $R\to1$, in its absence $R\to0$, and at a continuous transition $R$ becomes independent of $L$, so that the curves of different sizes cross. 

Each projective measurement of the system qubits yields one spin configuration of the prepared state, so the measured bit strings contain more information than their averages. We therefore also analyze the individual outcomes (snapshots), measured in the $Z$ basis with the IBM Quantum Runtime SamplerV2 primitive. From each shot with the ancilla in $|0_a\rangle$, we evaluate the magnetization $m=\frac{1}{L}\sum_j z_j$ and the $Z$ component of the staggered dimerization $d=\frac{1}{L}\sum_j(-1)^j z_jz_{j+1}$, with $z_j=\pm1$; the $X_jX_{j+1}$ part of $B_j$ is not accessible in a single $Z$-basis shot. Their joint distribution $P(m,d)$ shows, configuration by configuration, whether a state is ferromagnetic, dimerized, or composed of both types of configurations. Readout errors are mitigated with per-qubit confusion matrices obtained from calibration circuits, with all qubits prepared in $|0\rangle$ or $|1\rangle$, executed in the same job.

\section{Results}
\label{sec:results}
\subsection{Observation of the zFM--VBS crossover on IBMQ}
\begin{figure}[t]
    \centering
    \includegraphics[width=1\columnwidth]{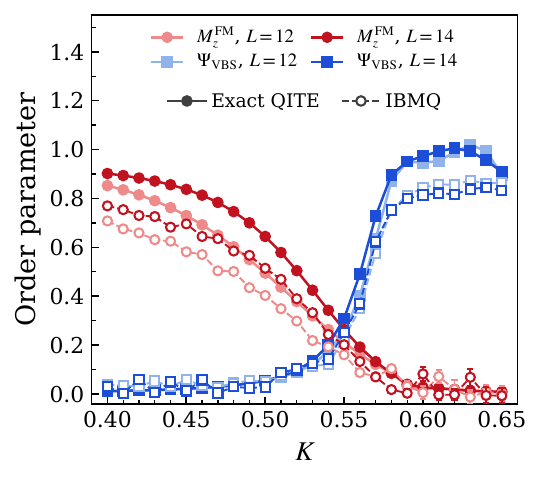}
    \caption{Order parameters $M_z^{\rm FM}$ and $\Psi_{\rm VBS}$ as functions of $K$ for $L=12$ and $14$.}
    \label{fig:phase_diagram}
\end{figure}
We first characterize the two phases and the transition between them through the competing order parameters, the magnetization $M_z^{\rm FM}$ and the dimerization $\Psi_{\rm VBS}$ defined in Eqs.~\ref{eq:Mz} and \ref{eq:PsiVBS}, respectively. Fig.~\ref{fig:phase_diagram} shows both as functions of $K$ for $L=12$ and $L=14$, comparing the IBMQ results with exact QITE. For weak $K$, the ferromagnetic NN coupling dominates and the spins align along $z$: $M_z^{\rm FM}$ is large and $\Psi_{\rm VBS}\approx0$, characteristic of the zFM phase. For strong $K$, the frustrating NNN coupling destroys the magnetic order and the spins pair up on alternating bonds: $M_z^{\rm FM}$ vanishes and $\Psi_{\rm VBS}$ approaches $1$, the value of a perfectly dimerized chain, characteristic of the VBS phase. In between, $M_z^{\rm FM}$ decreases and $\Psi_{\rm VBS}$ rises sharply for $K\approx0.54$--$0.58$; on these finite chains the transition appears as a smooth crossover. The IBMQ data reproduce this behavior for both system sizes. In the ordered regimes, the measured order parameters are smaller than the exact values; this reduction is caused by device noise and is analyzed in Sec.~\ref{sec:noise}. Since the order parameters of a finite chain vary smoothly with $K$, we determine the transition point from the crossing of the two order parameters and its dependence on $L$ in the next subsection.

\subsection{Finite-size extrapolation}
\label{sec:fss}

\begin{figure}
    \centering
    \includegraphics[width=1.0\columnwidth]{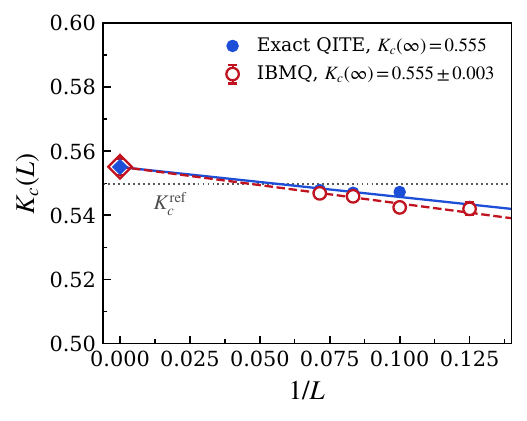}
    \caption{Order-parameter crossing $K_c(L)$ as a function of $1/L$. Lines are weighted linear fits $K_c(L)=K_c(\infty)+a/L$, and diamonds mark the extrapolated values. IBMQ error bars are propagated from the measurement errors by Monte Carlo sampling. The dotted line marks $K_c=0.5497$ from Ref.~\cite{Roberts2019}.}
    \label{fig:finite_size_extrapolation}
\end{figure}

We characterize the transition by the finite-size transition point $K_c(L)$, defined as the coupling at which the two order parameters cross, $M_z^{\rm FM}(K_c)=\Psi_{\rm VBS}(K_c)$, obtained by cubic interpolation in $K$. On a finite chain, both order parameters remain nonzero near the transition, so that $K_c(L)$ is only an estimate of the critical coupling; it converges to $K_c$ in the thermodynamic limit, where $M_z^{\rm FM}$ vanishes for $K>K_c$ and $\Psi_{\rm VBS}$ for $K<K_c$. Fig.~\ref{fig:finite_size_extrapolation} shows $K_c(L)$ for $L=8$--$14$ from exact QITE and IBMQ as a function of $1/L$. With increasing $L$, the crossing shifts to larger $K$ in both data sets, from $0.543$ for $L=8$ to $0.548$ for $L=14$ in exact QITE. The IBMQ crossing points closely follow the exact ones and lie slightly below them, by at most $0.005$. This small shift is expected: noise drives the state towards the maximally mixed state, in which both order parameters vanish, and since it reduces $M_z^{\rm FM}$ more strongly than $\Psi_{\rm VBS}$, the crossing moves to smaller $K$. The uncertainty of the IBMQ crossing points decreases with $L$, as the order parameters vary more steeply with $K$ for longer chains.
A weighted linear fit $K_c(L)=K_c(\infty)+a/L$ gives $K_c(\infty)=0.555\pm0.003$ for the IBMQ data, consistent within error bounds with the exact-QITE value $K_c(\infty)=0.555$. Although noise reduces the magnitude of both order parameters, its impact on the crossing point is minimal, allowing the hardware to capture the transition region closely mapped by the noiseless calculation for the same system sizes. Both extrapolations are also broadly consistent with the thermodynamic estimate $K_c=0.5497$ from finite-entanglement scaling~\cite{Roberts2019}. 
Having located the transition, we next examine the two phases on either side of it through their spin and dimer correlations.

\subsection{Spin and dimer correlations in real space}
\label{sec:results_corr}

\begin{figure}
    \centering
    \includegraphics[width=1.0\columnwidth]{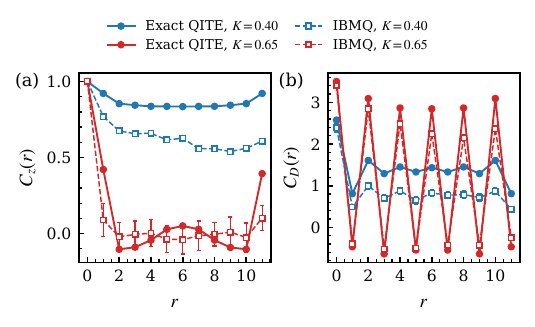}
    \caption{(a) Spin correlation $C_z(r)$ and (b) dimer correlation $C_D(r)$ for $L=12$.}
    \label{fig:corr}
\end{figure}

To characterize the two phases beyond the order parameters, we examine the real-space spin and dimer correlation functions at $K=0.40$ (zFM side) and $K=0.65$ (VBS side), shown in Fig.~\ref{fig:corr} for $L=12$. At $K=0.40$, the spin correlation $C_z(r)$ saturates at a finite value for all separations, the real-space signature of long-range zFM order. The IBMQ data remain finite across the entire chain; noise reduces their amplitude, but the long-range ferromagnetic correlation is clearly retained. At $K=0.65$, $C_z(r)$ vanishes beyond the nearest neighbors, signalling the loss of zFM order. Only the nearest-neighbor correlation remains finite ($r=1$ and $r=L-1$), because neighboring spins form pairs whose $z$ components are aligned. On hardware, the vanishing long-range correlation is reproduced, but this short-range correlation is not resolved.
The dimer correlation function in Fig.~\ref{fig:corr}(b) provides the complementary signature. At $K=0.65$, $C_D(r)$ shows a pronounced even--odd oscillation that persists over the entire chain, reflecting the long-range staggered bond pattern of the VBS phase. The oscillation is clearly resolved in the IBMQ data, with an amplitude reduced by noise; the dimer correlator is more susceptible to noise than the spin correlator because it involves Pauli strings of weight up to four. At $K=0.40$, both the exact and the IBMQ correlations show a nearly uniform background without a persistent even--odd structure, confirming the absence of dimerization in the zFM phase. These real-space signatures of the two phases are quantified in momentum space in the next subsection.

\subsection{Structure-factor signatures of zFM and VBS order}

\begin{figure}[t]
    \centering
    \includegraphics[width=\columnwidth]{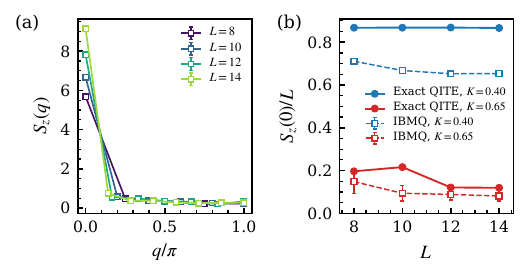}
    \caption{(a) Spin structure factor $S_z(q)$ measured on IBMQ at $K=0.40$ and (b) $S_z(0)/L$ as a function of $L$.}
    \label{fig:spin_sf}
\end{figure}

The spin and dimer structure factors provide the momentum-space signatures of the two phases. At $K=0.40$, the spin structure factor $S_z(q)$ measured on IBMQ shows a pronounced peak at $q=0$, as shown in Fig.~\ref{fig:spin_sf}(a), the signature of ferromagnetic correlations. Whether these correlations are long ranged is tested by $S_z(0)/L$, which remains finite in the thermodynamic limit only in the presence of long-range ferromagnetic order, as shown in Fig.~\ref{fig:spin_sf}(b). At $K=0.40$, $S_z(0)/L$ is large and nearly independent of $L$, whereas at $K=0.65$ it is small and decreases overall with increasing $L$. Although the IBMQ values are smaller than the exact ones, they show the same trends.
\begin{figure}[b]
    \centering
    \includegraphics[width=\columnwidth]{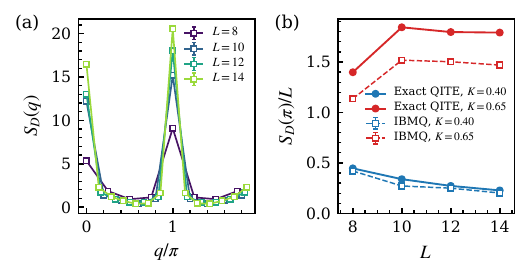}
    \caption{(a) Dimer structure factor $S_D(q)$ measured on IBMQ at $K=0.65$; the peak at $q=0$ arises from the uniform background $\langle B\rangle^2_{\mathcal P}$. (b) $S_D(\pi)/L$ as a function of $L$.}
    \label{fig:dimer_sf}
\end{figure}
At $K=0.65$, the dimer structure factor $S_D(q)$ measured on IBMQ shows a sharp peak at $q=\pi$, the ordering wave vector of the staggered bond pattern, for all system sizes, as shown in Fig.~\ref{fig:dimer_sf}(a). The additional peak at $q=0$ arises from the uniform background $\langle B\rangle^2_{\mathcal P}$ in $C_D(r)$ and is unrelated to dimer order. Correspondingly, $S_D(\pi)/L$ remains large and almost constant with $L$ at $K=0.65$, whereas at $K=0.40$ it is small and decreases with increasing $L$, as shown in Fig.~\ref{fig:dimer_sf}(b). The IBMQ data reproduce this behavior for all $L$; at $K=0.65$ their values are somewhat smaller than the exact ones, since the four-site dimer correlator involves multi-qubit Pauli strings that experience higher cumulative depolarization and error accumulation during hardware execution(see Noise Model in sec.~\ref{sec:noise}).
\begin{figure}[t]
    \centering
    \includegraphics[width=\columnwidth]{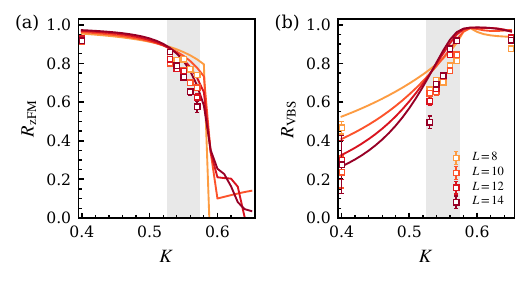}
    \caption{(a) $R_{\rm zFM}$ and (b) $R_{\rm VBS}$ as functions of $K$: exact QITE (lines) and IBMQ (symbols). The shaded region marks the transition region. $R_{\rm zFM}$ is not shown on hardware at $K=0.65$, where $S_z(q)$ has no peak.}
    \label{fig:ratio}
\end{figure}

\subsection{Correlation ratios}
\label{sec:ratio}

The order parameters and the structure factors per site characterize the two phases, but near the transition their size dependence is difficult to interpret, because noise reduces their magnitude and the reduction differs between system sizes. The correlation ratio defined in Eq.~\ref{eq:ratio} avoids this problem: as a ratio of two structure-factor values, it is insensitive to a uniform reduction of the correlations. It approaches $1$ for long-range order and $0$ for short-range correlations, so that the direction of its size dependence indicates, for each order separately, on which side of the transition a state lies.

Fig.~\ref{fig:ratio} shows $R_{\rm zFM}$ and $R_{\rm VBS}$ for $L=8$--$14$. Deep in the zFM phase ($K=0.40$), $R_{\rm zFM}$ is close to $1$ (above $0.9$ on hardware), whereas $R_{\rm VBS}$ is small and decreases with $L$, as expected for ferromagnetic order without dimerization; deep in the VBS phase ($K=0.65$), $R_{\rm VBS}$ approaches $1$. Within the critical region $0.53\lesssim K\lesssim0.57$, the two ratios behave in opposite ways: $R_{\rm zFM}$ falls and $R_{\rm VBS}$ rises with increasing $K$, and in the middle of this region they become comparable ($R_{\rm zFM}\approx R_{\rm VBS}\approx0.75$--$0.85$), i.e., the ferromagnetic and dimer correlations are equally long ranged. The change depends strongly on the system size. For the smallest chain ($L=8$), both ratios vary only slowly and remain high across the region, with $R_{\rm zFM}$ decreases from $0.84$ to $0.74$ and $R_{\rm VBS}$ increases from $0.66$ to $0.81$ between $K=0.53$ and $0.57$. For the largest chain ($L=14$), the evolution is much sharper: $R_{\rm zFM}$ drops steeply from $0.86$ to $0.58$ and $R_{\rm VBS}$ rises from $0.50$ to $0.92$. Consequently, the parameter window where both correlations are comparably long-ranged narrows significantly with increasing system size. The IBMQ data reproduce these trends and the reversal of both ratios, although their values lie below the exact ones, since noise does not attenuate the correlations uniformly at all distances.

\subsection{Snapshot distributions}
\label{sec:snapshots}
\begin{figure}[t]
    \centering
    \includegraphics[width=\columnwidth]{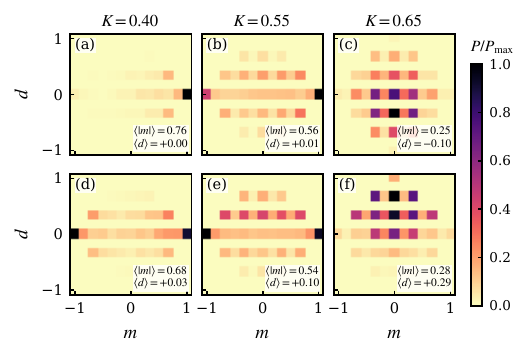}
    \caption{Distribution $P(m,d)$ of the magnetization $m$ and dimerization $d$ of the post-selected $Z$-basis snapshots for $L=12$ on IBMQ (readout-error mitigated): (a)--(c) $h_z$-pinned and (d)--(f) $h_d$-pinned circuits. Colours show $P/P_{\max}$ in each panel; boxes give $\langle|m|\rangle$ and $\langle d\rangle$.}
    \label{fig:snapshots}
\end{figure}

The correlation functions and their ratios are averages over many measurements and do not reveal the individual configurations from which these averages arise. Such averages cannot distinguish, for example, an ordered state that is a superposition of oppositely ordered configurations from a disordered one, or a state in which each configuration is partially ordered from one composed of distinct ordered configurations. The $Z$-basis snapshots introduced in Sec.~\ref{sec:methods} resolve this, since each post-selected shot yields one spin configuration. Fig.~\ref{fig:snapshots} shows the distribution $P(m,d)$ of the magnetization and dimerization of these shots for $L=12$, at $K=0.40$ (zFM), $0.55$ (near the critical region), and $0.65$ (VBS).

At $K=0.40$, about half of the shots are fully polarized, at $m=+1$ in Fig.~\ref{fig:snapshots}(a) and at both $m=+1$ and $m=-1$ in Fig.~\ref{fig:snapshots}(d). In the latter case the average magnetization vanishes, although every shot is ferromagnetic; the snapshots thus reveal ferromagnetic order that the average conceals, and the boxes therefore give $\langle|m|\rangle$. At $K=0.55$, fully polarized configurations coexist with weakly magnetized and partially dimerized ones, showing that the two orders compete shot by shot rather than through a uniform intermediate state. At $K=0.65$, fully polarized shots have disappeared; the weight is centred at $m\approx0$ and shifted to $d>0$, as shown in Fig.~\ref{fig:snapshots}(f), reflecting the staggered dimer order of the VBS phase.

\subsection{Noise model}
\label{sec:noise}

The order parameters and correlations measured on IBMQ follow the exact trends but with reduced amplitudes. Two sources can cause such deviations: the optimized circuit approximates the target state only with finite fidelity (compression error), and the hardware executes this circuit imperfectly (device noise). To separate them, we simulate the executed circuits with a noise model of \texttt{ibm\_kingston}, built from the device calibration closest to each hardware job, using Qiskit Aer~\cite{Qiskit2024}. For every physical qubit and CZ pair used, the model includes the calibrated gate errors as depolarizing channels~\cite{Magesan2011} and thermal relaxation during each gate, set by the calibrated $T_1$, $T_2$ and gate durations~\cite{NielsenChuang2010}; readout errors are omitted, which corresponds to ideal TREX mitigation~\cite{vandenBerg2022}. The circuits are evolved as density matrices on the same physical qubits as on hardware. All error rates are scaled by a common factor $\lambda$ (see Appendix~\ref{app:noise}): $\lambda=0$ gives the noiseless circuit and thus isolates the compression error, $\lambda=1$ corresponds to the calibrated device, and the fitted $\lambda$ gives the actual noise level of a run. We first apply this model to the order parameters, which separates the compression error from the device noise and determines $\lambda$. We then show that the projection probability $p_0$ controls how strongly noise affects a post-selected observable, and finally discuss the features of the correlation functions that follow from the structure of the observables and circuits.

\begin{figure*}[t]
    \centering
    \includegraphics[width=\textwidth]{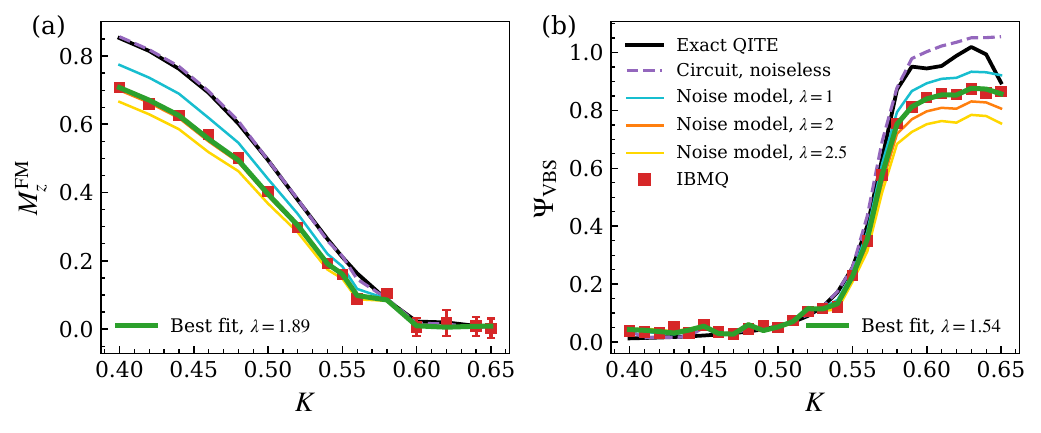}
    \caption{Device-noise simulation of (a) $M_z^{\rm FM}$ and (b) $\Psi_{\rm VBS}$ for $L=12$, compared with exact QITE and the IBMQ data of Fig.~\ref{fig:phase_diagram}. The noise scale $\lambda$ multiplies all calibrated error rates; the best-fit values are given in each panel.}
    \label{fig:noise_model}
\end{figure*}

\emph{\textbf{Order parameters.}} Fig.~\ref{fig:noise_model} compares the simulations with the hardware data for $L=12$. The compression error is small: the noiseless circuit follows exact QITE closely, except deep in the VBS phase, where it overestimates $\Psi_{\rm VBS}$, by up to $0.16$ at $K=0.65$, because the optimized state is more strongly dimerized than the exact one, mainly on the bond $(L-1,0)$ that closes the ring and is generated by the open-chain ansatz only through its full depth. The deviations of the hardware data are therefore dominated by device noise. The calibrated error rates ($\lambda=1$) account for only part of these deviations, whereas a single scale factor per run reproduces the hardware data in both sectors, including the sharp rise of $\Psi_{\rm VBS}$ across the transition and its reduced plateau in the VBS phase. The fitted values, $\lambda=1.89$ for $M_z^{\rm FM}$ and $1.54$ for $\Psi_{\rm VBS}$ at $L=12$, and $1.34$--$1.95$ for all runs with $L=8$--$12$ (Table~\ref{tab:lambda}), vary between runs recorded on different days and qubits but show no systematic dependence on system size. 
The effective error rates during execution thus exceed the calibrated ones by a factor of $1.3\text{--}2$. This discrepancy arises from environmental and operational factors absent from standard calibration protocols (which typically evaluate gates in isolation [36]), including inter-qubit crosstalk, coherent (systematic) gate errors, and temporal drift of device characteristics between calibration runs and execution.

\begin{table}[t]
\caption{Best-fit noise scale $\lambda$ ($1\sigma$ interval) and $\chi^2$ per data point at the best fit and at $\lambda=1$.}
\label{tab:lambda}
\centering
\renewcommand{\arraystretch}{1.35}\setlength{\tabcolsep}{5pt}
\begin{tabular}{|c|c|c|c|c|}
\hline
\multirow{2}{*}{$L$} & \multirow{2}{*}{sector} & \multirow{2}{*}{$\lambda$} & \multicolumn{2}{c|}{$\chi^2/N$} \\ \cline{4-5}
 & & & best fit & $\lambda=1$ \\
\hline\hline
\multirow{2}{*}{8}  & zFM & 1.75 (1.68--1.82) & 1.43 & 6.0 \\ \cline{2-5}
                    & VBS & 1.68 (1.62--1.74) & 1.51 & 6.5 \\ \hline
\multirow{2}{*}{10} & zFM & 1.95 (1.91--1.98) & 3.11 & 34.0 \\ \cline{2-5}
                    & VBS & 1.34 (1.32--1.36) & 1.12 & 10.8 \\ \hline
\multirow{2}{*}{12} & zFM & 1.89 (1.86--1.92) & 1.21 & 47.3 \\ \cline{2-5}
                    & VBS & 1.54 (1.51--1.57) & 0.51 & 13.3 \\ \hline
\end{tabular}
\end{table}

\begin{table}[b]
\caption{Projection probability $p_0$ for $L=12$: ideal circuit, noise model with fitted $\lambda$, and IBMQ.}
\label{tab:p0}
\centering
\renewcommand{\arraystretch}{1.35}\setlength{\tabcolsep}{5pt}
\begin{tabular}{|c|c|c|c|c|}
\hline
circuit & $K$ & ideal & noise model & IBMQ \\ \hline \hline
\multirow{2}{*}{zFM} & 0.40 & 0.763 & 0.735 & 0.715 \\ \cline{2-5}
                     & 0.65 & 0.016 & 0.056 & 0.088 \\ \hline
\multirow{2}{*}{dimer} & 0.40 & 0.160 & 0.199 & 0.198 \\ \cline{2-5}
                     & 0.65 & 0.665 & 0.651 & 0.642 \\ \hline
\end{tabular}
\end{table}

\emph{\textbf{Projection probability.}} Not all observables are equally affected by noise, mainly because of the post-selection. Noise redistributes weight between the two ancilla branches, so that the measured projection probability also contains erroneous runs. In a global depolarizing approximation with circuit fidelity $F_{\rm c}$~\cite{Urbanek2021}, these add $(1-F_{\rm c})/2$ to the measured $p_0$ but nothing to $\langle P_0O\rangle$ for a traceless observable $O$, which reduces the post-selected signal by the factor (see Appendix~\ref{app:dilution})
\begin{equation}
\langle O\rangle_{\mathcal P}^{\rm noisy} \simeq s\,\langle O\rangle_{\mathcal P},
\qquad
s=\frac{F_{\rm c}\, p_0}{F_{\rm c}\, p_0 + (1-F_{\rm c})/2}.
\label{eq:dilution}
\end{equation}
The reduction is therefore strongest when the ideal projection probability $p_0$ is small, as shown in  Table~\ref{tab:p0}. Where $p_0$ is large, noise changes it only slightly, conversely where it is small, the measured $p_0$ increases several-fold, from $0.016$ to $0.088$ for the zFM circuit at $K=0.65$, so that most of the kept shots are erroneous. This is why the nearest-neighbor spin correlation $C_z(1)\approx0.4$ of the exact state at $K=0.65$ is not resolved on hardware, as shown in Fig.~\ref{fig:corr}(a). On hardware, the attenuation is weaker than predicted by Eq.~\ref{eq:dilution}, and compiling the circuits against the exact projection probability keeps $p_0$ large (see Appendix~\ref{app:p0_draft}).\\

\emph{\textbf{Correlation functions.}} 
\afterpage{\begin{figure}[t]
    \centering
    \includegraphics[width=\columnwidth]{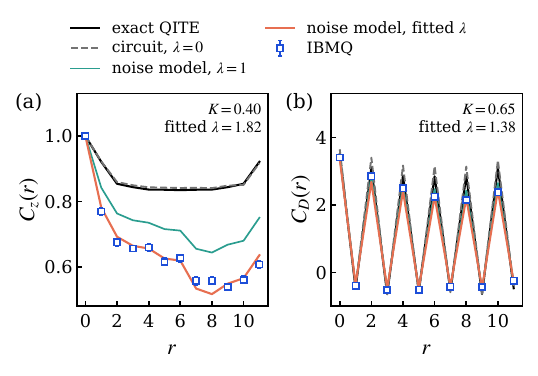}
    \caption{Device-noise simulation of (a) $C_z(r)$ at $K=0.40$ and (b) $C_D(r)$ at $K=0.65$ for $L=12$, compared with exact QITE and the IBMQ data of Fig.~\ref{fig:corr}.}
    \label{fig:noise_corr}
\end{figure}}
The noise model also accounts for the correlation functions, as shown in Fig.~\ref{fig:noise_corr}. The spin correlations retain about $0.72$ of their exact values on average and the dimer correlations about $0.84$, consistent with the larger noise scale fitted to the spin run ($\lambda=1.82$) than to the dimer run ($\lambda=1.38$). In both cases the attenuation grows with distance: because the ansatz is an open chain, the correlation between the reference site and site $r=L-1$, its neighbor on the periodic ring, is mediated across the entire hardware chain, which breaks the $r\to L-r$ symmetry of the exact correlations. With $\lambda$ fitted to each run, the simulations reproduce the measured correlations including this asymmetry. Improving the hardware results therefore requires larger projection probabilities and error mitigation beyond readout correction~\cite{Temme2017,Kim2023}.

\section{Conclusion}
\label{sec:conclusion}

We have studied the transition between a $z$-polarized ferromagnet and a valence-bond solid in a frustrated spin-$1/2$ chain, the one-dimensional realization of a deconfined quantum critical point, on a digital quantum processor. The ground states were prepared by ancilla-projected QITE, whose exact target states were compressed into shallow parametrized brickwork quantum circuits and executed on \texttt{ibm\_kingston} for chains of up to $L=14$ sites. The hardware resolves both ordered phases. The order parameters display the crossover between them; in real space, the zFM phase shows long-range ferromagnetic correlations without bond modulation and the VBS phase long-range staggered dimer correlations; and in momentum space, the spin and dimer structure factors peak at $q=0$ and $q=\pi$, respectively, with the size dependence expected for long-range order. The finite-size extrapolation of the order-parameter crossing gives $K_c = 0.555 \pm 0.003$, consistent with the thermodynamic value $0.5497$~\cite{Roberts2019}. The correlation ratios of the two orders, which are largely insensitive to the noise-induced attenuation, reproduce on hardware the reversal of their size dependence across the transition. Furthermore, The measured snapshots show how the spin configurations change from ferromagnetic to dimerized across the transition. We also study noise model based on the device calibration shows that the deviations from exact QITE are dominated by hardware noise, and it reproduces the data in both phases when all error rates are scaled by a single factor $\lambda\approx1.3$--$2$ per run. The main limitation of the present framework is the small projection probability of some circuits, which amplifies the effect of noise; compiling the circuits against the exact projection probability keeps $p_{0}$ large and gives well-resolved VBS data for all chain lengths. Resolving the critical properties themselves, such as the equal scaling of the two order parameters, will require larger systems, error mitigation beyond readout correction~\cite{Temme2017,Kim2023}. Our approach, which combines exact target states, tensor-network circuit compression, and post-selected execution, thus offers a practical way to prepare and study competing-order ground states on current quantum hardware. A natural next step is to construct the target states from matrix-product-state calculations~\cite{Schollwock2011}, so that the same circuit compression and post-selected execution on IBMQ can reach chain lengths beyond the reach of exact classical computation.

\begin{acknowledgments}
We acknowledge the use of IBM Open Plan Quantum services for this work. MKG thanks Prof. Pochung Chen for fruitful discussion. The computational work was performed on resources of the National Supercomputing Centre, Singapore. The authors used Claude (Anthropic) to improve the presentation of the figures and for language editing. All suggestions were reviewed and approved by the authors. PS and MKG acknowledge the support of the Singapore Ministry of Education Academic Research Fund’s Tier 3 Grant MOE-MOET32023-0003 (PS) and Tier 2 Grant MOET2EP50123-0021 (MKG).
\end{acknowledgments}

\appendix
\section{Device-noise simulations}
\label{app:noise}

\subsection{Noise channels}
\label{app:channels}
A noise channel $\mathcal{E}$ describes how an imperfect operation changes the density matrix of the qubits, $\rho\to\mathcal{E}(\rho)$~\cite{NielsenChuang2010}. In our model, every native gate (\texttt{sx}, \texttt{x}, or CZ) is followed by a channel with two contributions: thermal relaxation of the qubits during the gate, and a residual gate error that is not explained by relaxation. The virtual $R_z$ gates are error-free.

\emph{Thermal relaxation.} During a gate $g$ of duration $t_g$, each qubit $q$ of the gate relaxes towards $|0\rangle$ with its energy-relaxation time $T_1^{(q)}$ and loses phase coherence with its dephasing time $T_2^{(q)}$. In terms of the Bloch-vector components of qubit $q$,
\begin{align}
\langle X\rangle&\to e^{-t_g/T_2^{(q)}}\langle X\rangle,\qquad
\langle Y\rangle\to e^{-t_g/T_2^{(q)}}\langle Y\rangle,\nonumber\\
\langle Z\rangle&\to e^{-t_g/T_1^{(q)}}\langle Z\rangle+1-e^{-t_g/T_1^{(q)}}.
\end{align}

\emph{Residual gate error.} The remaining error is modelled by a depolarizing channel, which with probability $\gamma_g$ replaces the state of the gate qubits by the maximally mixed state,
\begin{equation}
\mathcal{D}_{\gamma_g}(\rho)=(1-\gamma_g)\,\rho+\gamma_g\,\frac{I_d}{d}\otimes\mathrm{Tr}_{g}\,\rho ,
\end{equation}
where $d=2$ ($d=4$) for single-qubit (two-qubit) gates, $I_d$ is the identity on the gate qubits, and $\mathrm{Tr}_{g}$ is the partial trace over them. The noise channel following gate $g$ is $\mathcal{E}_g=\mathcal{T}_g\circ\mathcal{D}_{\gamma_g}$, where $\mathcal{T}_g$ denotes thermal relaxation of all qubits of the gate for the duration $t_g$.

\emph{Matching the calibrated gate error.} The gate error $\varepsilon_g$ reported in the calibration is the average gate infidelity obtained from randomized benchmarking~\cite{Magesan2011}. To combine errors from different sources, it is convenient to use the process infidelity $\eta=1-F_{\rm pro}$, where $F_{\rm pro}$ is the process fidelity. It is related to the average gate infidelity $\varepsilon$ of the same channel by $\eta=(d+1)\varepsilon/d$~\cite{Horodecki1999,Nielsen2002} and, for small errors, is additive. The process infidelity of thermal relaxation of qubit $q$ during gate $g$ is
\begin{equation}
\eta^{\rm th}_{q}=\frac{3-e^{-t_g/T_1^{(q)}}-2e^{-t_g/T_2^{(q)}}}{4},
\end{equation}
and that of the depolarizing channel is $\eta^{\rm dep}_g=\gamma_g\,(1-1/d^2)$. The depolarizing probability is fixed by requiring that the total error equals the calibrated one,
\begin{equation}
\frac{d+1}{d}\,\varepsilon_g=\sum_{q\in g}\eta^{\rm th}_q+\gamma_g\Big(1-\frac{1}{d^2}\Big),
\label{eq:pg}
\end{equation}
i.e., $\gamma_g$ accounts for the part of the calibrated error that is not explained by relaxation ($\gamma_g=0$ if relaxation alone exceeds $\varepsilon_g$). In the simulations, Eq.~\ref{eq:pg} is solved in its exact multiplicative form, $\gamma_g=(F^{\rm th}_{g}-F^{\rm cal}_{g})/(F^{\rm th}_{g}-1/d^2)$ with $F^{\rm th}_{g}=\prod_{q\in g}(1-\eta^{\rm th}_q)$ and $F^{\rm cal}_{g}=1-(d+1)\varepsilon_g/d$ the process fidelities of the relaxation channel and of the calibrated gate, as implemented in Qiskit Aer~\cite{Qiskit2024}; for the error rates considered here, both forms agree to better than $1\%$.

All parameters ($\varepsilon_g$, $t_g$, $T_1^{(q)}$, $T_2^{(q)}$) are taken individually for each physical qubit and each CZ pair from the \texttt{ibm\_kingston} calibration closest to the submission time of the respective hardware job. Readout errors are not included, which corresponds to ideal TREX readout mitigation~\cite{vandenBerg2022}. Since the hardware circuits are Pauli-twirled~\cite{Wallman2016}, the gate errors act effectively as stochastic Pauli noise, consistent with the incoherent channels used here.

\subsection{Noise scaling}
To account for noise not contained in the calibration data, all error rates are scaled by a common factor $\lambda$,
\begin{equation}
\varepsilon_g\to\lambda\,\varepsilon_g,\qquad T_{1,2}^{(q)}\to T_{1,2}^{(q)}/\lambda .
\end{equation}
For weak noise, $\eta^{\rm th}_q\simeq t_g\,\big(1/T_1^{(q)}+2/T_2^{(q)}\big)/4$ and $\gamma_g$ are both linear in the error rates, so that the infidelity of every gate is multiplied by $\lambda$. The value $\lambda=1$ corresponds to the calibrated device, and $\lambda=0$ to the noiseless circuit.

\subsection{Simulation and fit}
\begin{figure}[t]
    \centering
    \includegraphics[width=\columnwidth]{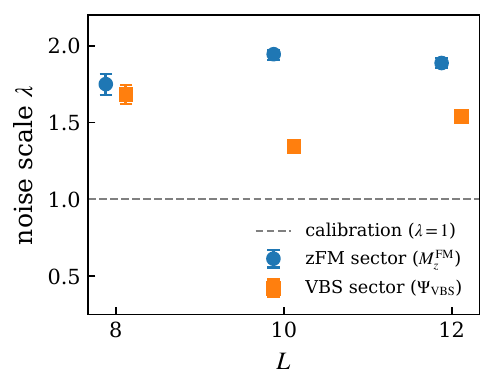}
    \caption{Best-fit noise scale $\lambda$ as a function of system size for the zFM ($M_z^{\rm FM}$) and VBS ($\Psi_{\rm VBS}$) sectors. Error bars mark the $1\sigma$ intervals of the fit; the dashed line indicates the calibrated device ($\lambda=1$).}
    \label{fig:lambda_vs_L}
\end{figure}

The executed circuit $U=g_N\cdots g_1$, compiled to the native gates on the physical qubits used on hardware, is evolved exactly in the density-matrix formalism,
\begin{equation}
\rho_\lambda=\big(\mathcal{E}_{g_N}\circ\mathcal{U}_{g_N}\circ\cdots\circ\mathcal{E}_{g_1}\circ\mathcal{U}_{g_1}\big)(\rho_{\rm in}),
\end{equation}
with $\mathcal{U}_g(\rho)=g\rho g^\dagger$, the noise channels $\mathcal{E}_g$ evaluated at scale $\lambda$, and $\rho_{\rm in}=|0\rangle\langle0|^{\otimes(L+1)}$,
and the observables are evaluated with the same ancilla-projected estimator as on hardware,
\begin{equation}
\langle O\rangle_{\mathcal{P}}(\lambda)=\frac{\mathrm{Tr}[P_0\,O\,\rho_\lambda]}{\mathrm{Tr}[P_0\,\rho_\lambda]} .
\end{equation}
The noise scale is determined separately for each sector and system size by minimizing
\begin{equation}
\chi^2(\lambda)=\sum_{K}\left[\frac{O_{\rm IBMQ}(K)-\langle O\rangle_{\mathcal{P}}(\lambda;K)}{\sigma_{\rm IBMQ}(K)}\right]^2,
\end{equation}
where the sum runs over the $N$ simulated couplings $K$, $\sigma_{\rm IBMQ}$ are the hardware error bars, and $\langle O\rangle_{\mathcal{P}}(\lambda;K)$ is interpolated linearly between the simulated values of $\lambda$. The $1\sigma$ interval is defined by $\chi^2\le\chi^2_{\rm min}+1$. The curves at the best-fit values were subsequently simulated directly and reproduce the interpolated $\chi^2$. The results are summarized in Table~\ref{tab:lambda} and Fig.~\ref{fig:lambda_vs_L}. The same fit applied to the $L=12$ correlation runs (separate jobs, $K=0.40$ and $0.65$, all distances $r\geq1$) gives $\lambda=1.82$ ($1.80$--$1.85$) for the spin run and $1.38$ ($1.36$--$1.41$) for the dimer run, within the range of the order-parameter runs.

\begin{figure*}[t]
    \centering
    \includegraphics[width=\textwidth]{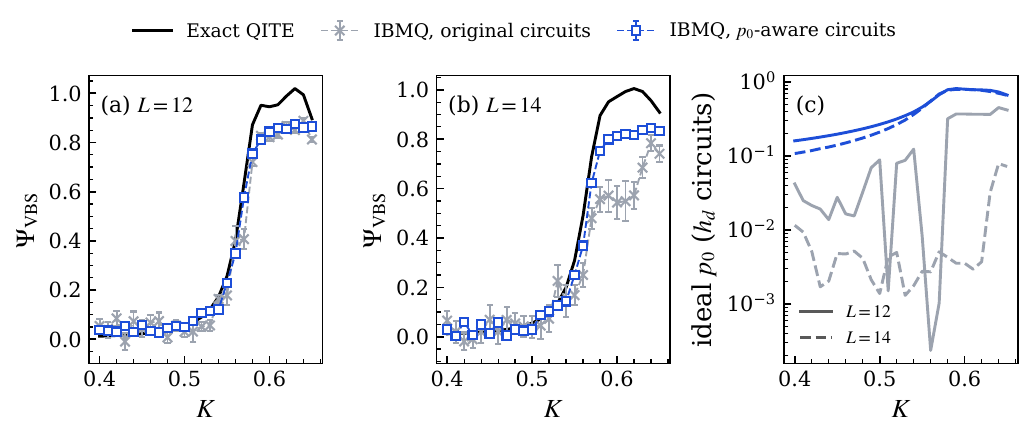}
    \caption{$p_0$-aware compilation. VBS order parameter for (a) $L=12$ and (b) $L=14$: exact QITE (line), IBMQ with the original circuits (crosses) and with the $p_0$-aware circuits (open squares). (c) Ideal projection probability of the original (grey) and $p_0$-aware (blue) circuits for $L=12$ (solid) and $L=14$ (dashed).}
    \label{fig:p0_aware}
\end{figure*}

\subsection{Post-selection dilution}
\label{app:dilution}
To make the role of $p_0$ explicit, approximate the effect of noise on the full $(L+1)$-qubit state by a global depolarizing channel with circuit fidelity $F_{\rm c}$,
\begin{equation}
\rho_{\rm noisy}\simeq F_{\rm c}\,\rho_{\rm ideal}+(1-F_{\rm c})\,\frac{I}{2^{L+1}},
\end{equation}
where $\rho_{\rm ideal}=\rho_{\lambda=0}$ is the noiseless output state. Since $\mathrm{Tr}\,P_0=2^{L}$, the projection probability becomes
\begin{equation}
\langle P_0\rangle_{\rm noisy}=F_{\rm c}\,p_0+\frac{1-F_{\rm c}}{2},\qquad p_0=\mathrm{Tr}[P_0\rho_{\rm ideal}].
\end{equation}
For a traceless system observable $O$, the operator $P_0O=\tfrac12(O+Z_aO)$, with $Z_a$ the Pauli-$Z$ operator of the ancilla, is also traceless, so the mixed component does not contribute to the numerator,
\begin{equation}
\langle P_0O\rangle_{\rm noisy}=F_{\rm c}\,\mathrm{Tr}[P_0O\rho_{\rm ideal}]=F_{\rm c}\,p_0\,\langle O\rangle_{\mathcal{P}} .
\end{equation}
Taking the ratio yields Eq.~\ref{eq:dilution},
\begin{equation}
\langle O\rangle_{\mathcal{P}}^{\rm noisy}=s\,\langle O\rangle_{\mathcal{P}},\qquad s=\frac{F_{\rm c}\,p_0}{F_{\rm c}\,p_0+(1-F_{\rm c})/2}.
\end{equation}
Noise thus increases the measured projection probability by $(1-F_{\rm c})(1/2-p_0)$ while reducing the numerator by the factor $F_{\rm c}$; for $p_0\ll(1-F_{\rm c})/2$, $s\approx 2F_{\rm c}\,p_0/(1-F_{\rm c})\ll1$ and the post-selected signal is lost. The global depolarizing form is a simplification used to expose the mechanism; it overestimates the attenuation at small $p_0$ (Sec.~\ref{sec:noise}), and the quantitative comparisons in the main text are based on the full noise model described above.

\section{Projection probability and $p_0$-aware compilation}
\label{app:p0_draft}

\begin{figure}[h]
    \centering
    \includegraphics[width=\columnwidth]{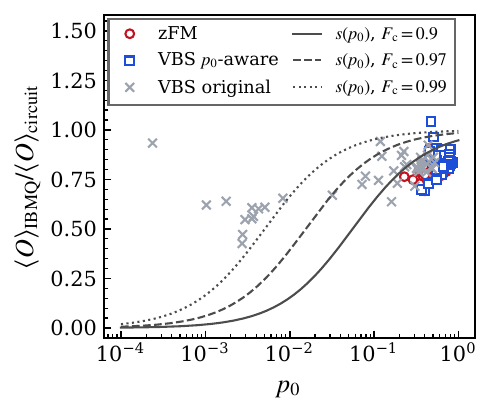}
    \caption{Fraction of the noiseless-circuit signal retained on hardware, $\langle O\rangle_{\rm IBMQ}/\langle O\rangle_{\rm circuit}$, versus the ideal projection probability $p_0$, for all runs and couplings with $|\langle O\rangle_{\rm circuit}|>0.15$. Lines: estimate of Eq.~\ref{eq:dilution} for three values of $F_{\rm c}$.}
    \label{fig:attenuation}
\end{figure}

\begin{figure*}[t]
    \centering
    \includegraphics[width=\textwidth]{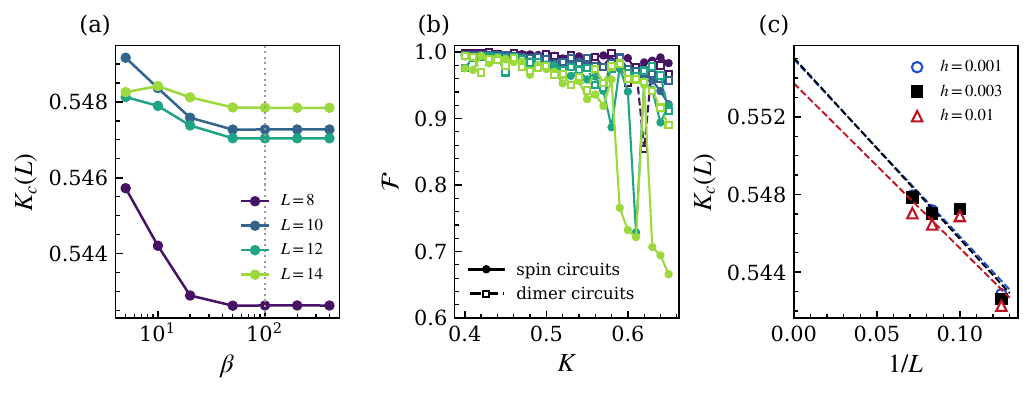}
    \caption{(a) Crossing point $K_c(L)$ of the exact-QITE order parameters as a function of $\beta$; the dotted line marks $\beta=100$. (b) Fidelity $\mathcal{F}$ of the spin and dimer circuits with their targets. (c) $K_c(L)$ as a function of $1/L$ for three pinning-field strengths, with linear fits.}
    \label{fig:validation}
\end{figure*}

The dimer (VBS) circuits were compiled twice for the same target states. In the original compilation, the reference projection probability in the penalty term of Eq.~\ref{eq:cost} was set too low, so that the optimizer accepted circuits with very small $p_0$; the $p_0$-aware compilation uses the exact QITE value. The spin (zFM) circuits were compiled against the correct reference from the start, so that their projection probability equals that of the exact target. Both sets of dimer circuits were executed on \texttt{ibm\_kingston}, which provides a controlled test of the role of $p_0$ at fixed target states.

Fig.~\ref{fig:p0_aware} compares the two compilations. The original circuits have $p_0\approx10^{-3}$--$10^{-1}$ over most of the range and $p_0\approx2\times10^{-4}$--$10^{-2}$ near the transition, whereas the $p_0$-aware circuits keep $p_0\geq0.1$ for all $K$, as shown in Fig.~\ref{fig:p0_aware}(c). For $L=12$, both give the same VBS plateau for $K\geq0.58$, $0.87$ and $0.88$ of the exact value, but the median error bar is halved ($0.023\to0.010$). For $L=14$, where the original $p_0$ is smallest, the $p_0$-aware circuits raise the plateau from $0.65$ to $0.85$ of the exact value and reduce the errors five-fold ($0.046\to0.009$). Since the target states are identical, the improvement is entirely due to the projection probability.

Fig.~\ref{fig:attenuation} tests whether this behavior is general. Each symbol is one hardware measurement of an order parameter, divided by the value of the noiseless circuit, i.e., the fraction of the signal that survives on hardware, plotted against the ideal $p_0$ of the circuit. Only couplings at which the order parameter is clearly nonzero, $|\langle O\rangle_{\rm circuit}|>0.15$, are included: $K=0.40$--$0.56$ for $M_z^{\rm FM}$ and $K=0.54$--$0.65$ for $\Psi_{\rm VBS}$, for $L=8$--$14$. For the spin circuits and the $p_0$-aware dimer circuits, $p_0\gtrsim0.2$, and about $0.82$ of the signal survives, nearly independently of the circuit and system size. Only the original dimer circuits reach small $p_0$; there, less survives, about $0.6$ for $p_0\sim10^{-3}$, with larger scatter. The lines show the estimate of Eq.~\ref{eq:dilution} for circuit fidelities $F_{\rm c}=0.9$, $0.97$, and $0.99$: each curve approaches $1$ for large $p_0$ and drops to zero for small $p_0$, reaching one half at $p_0=(1-F_{\rm c})/2F_{\rm c}$. The estimate thus reproduces the trend but predicts an almost complete loss of the signal at $p_0\sim10^{-3}$, where the hardware still retains about $0.6$. The real loss is smaller because hardware noise acts on a few qubits at a time rather than scrambling the whole state, so that fewer erroneous shots pass the ancilla filter; the full device-noise model (see Sec.~\ref{sec:noise}) reproduces the data for both compilations. The main cost of a small $p_0$ is therefore statistical: since only a fraction $p_0$ of the shots is kept, the error bars grow as $1/p_0$, as for the original circuits with $L=14$ in Fig.~\ref{fig:p0_aware}(b). Keeping $p_0$ large, by compiling against the exact projection probability and choosing a suitable initial state, is thus a simple and effective way to improve post-selected QITE on hardware.

\section{Validation of the targets, circuits, and pinning fields}
\label{app:validation}

Here we verify the imaginary-time convergence of the targets, the fidelity of the circuits, and the choice of the pinning-field strength.

\emph{Imaginary-time convergence.} The exact QITE states were evaluated for $\beta=5$--$400$ from the same initial states as the targets, as shown in Fig.~\ref{fig:validation}(a). The crossing points $K_c(L)$ converge from $\beta\approx50$ on; at $\beta=100$ they differ from their $\beta=400$ values by less than $10^{-5}$. The order parameters at $\beta=100$ deviate from those at $\beta=400$ by at most $1.5\times10^{-3}$ ($\Psi_{\rm VBS}$ for $L=8$) and by at most $1.5\times10^{-4}$ for $L\geq12$. The only exception is $M_z^{\rm FM}$ for $L=8$ deep in the VBS phase ($K\geq0.59$), where it is small and converges more slowly (deviation up to $0.04$).

\emph{Circuit fidelity.} Fig.~\ref{fig:validation}(b) shows the fidelity $\mathcal{F}=|\langle\psi_{\mathcal P}|\psi^{(0)}_{\bm\theta}\rangle|^2$ of the executed spin and dimer circuits with their targets. The median $\mathcal{F}$ decreases from $0.99$ for $L=8$ to $0.96$ (spin) and $0.97$ (dimer) for $L=14$, since the circuit depth grows more slowly than the entanglement of the target. In the transition region ($0.53\leq K\leq0.57$), $\mathcal{F}>0.91$ for all circuits. The lowest values, $0.67$--$0.73$ for $L=12$ and $14$, occur for the spin circuits deep in the VBS phase ($K\gtrsim0.60$). These circuits enter only $M_z^{\rm FM}$ in this region and the spin correlations at $K=0.65$, and in both cases they reproduce the absence of ferromagnetic order of the target: $M_z^{\rm FM}$ of the noiseless circuit deviates from the exact value by at most $0.08$, and $S_z(0)/L$ at $K=0.65$ by at most $0.02$. The results are therefore not affected.

\emph{Pinning field.} Fig.~\ref{fig:validation}(c) shows $K_c(L)$ for the pinning-field strengths $h=0.001$, $0.003$, and $0.01$, applied as $h_z$ and $h_d$ in the respective sectors. The crossing points change by less than $0.001$, and the extrapolated values are $K_c(\infty)=0.5550$, $0.5550$, and $0.5537$. The field strength $h=0.003$ used throughout is therefore large enough to select the broken-symmetry states for $L\geq10$ and small enough not to shift the transition.

\FloatBarrier
\bibliography{references}

\end{document}